\documentclass[twocolumn,apl,amsmath,amssymb,showpacs,superscriptaddress]{revtex4-2}
\usepackage{epsf} 
\usepackage{graphicx}
\usepackage{color}
\usepackage{soul}
\usepackage{gensymb}
\usepackage{sidecap}
\usepackage{amsmath}
\usepackage{mathtools}
\usepackage{float}
 \usepackage{multirow}
\usepackage[hidelinks,colorlinks=true,linkcolor=blue,citecolor=blue]{hyperref}
\DeclareUnicodeCharacter{2212}{\textendash}
\usepackage{tabularx}

\begin{document}
\title{Lattice dynamics and structural correlations in antiferromagnet SrLaCoNbO$_6$ double perovskite}

\author{Ajay Kumar}
\email{Present address: Ames National Laboratory, U.S. Department of Energy, Iowa State University, Ames, Iowa 50011, USA}
\affiliation{Department of Physics, Indian Institute of Technology Delhi, Hauz Khas, New Delhi-110016, India}

\author{Clemens Ulrich}
\affiliation{School of Physics, The University of New South Wales (UNSW), Kensington, 2052 New South Wales, Australia}

\author{Rajendra S. Dhaka}
\email{rsdhaka@physics.iitd.ac.in}
\affiliation{Department of Physics, Indian Institute of Technology Delhi, Hauz Khas, New Delhi-110016, India}

\date{\today}      

\begin{abstract}
We report a detailed temperature-dependent Raman spectroscopy and Co $K$-edge extended x-ray absorption fine structure (EXAFS) study of the insulating antiferromagnetic double perovskite SrLaCoNbO$_6$. Interestingly, pronounced changes in the phonon frequencies, linewidths, and intensities are observed around 60~K, well above the long-range antiferromagnetic ordering temperature, $T_{\rm N} \approx 15$~K. These anomalies indicate a modification of the lattice dynamics that may involve local structural distortions, short-range magnetic correlations, or an interplay between them. Additional changes in both high- and low-energy Raman modes are detected near 160--180~K via variations in phonon intensities and linewidths, together with changes in the Fano asymmetry of selected low-energy modes. The asymmetric line shapes of these modes are well described phenomenologically by Fano profiles and indicate coupling of the discrete phonons to an underlying continuum, whose possible microscopic origins are discussed. Moreover, the EXAFS analysis reveals correlated changes in bond distances and Debye--Waller factors around the Co ions near 60~K and 160~K, indicating subtle local structural distortions at temperatures corresponding to the Raman anomalies. The combined Raman and EXAFS results therefore demonstrate a close correlation between anomalous lattice dynamics and local structural evolution in SrLaCoNbO$_6$. 

\end{abstract}

\maketitle

\section{\noindent ~Introduction}

The cobalt based materials providing a versatile platform for investigating the interplay among lattice, spin, electronic, and orbital degrees of freedom have been extensively studied in last two decades due to their unconventional structural and magnetic responses over a broad temperature range \cite{Liu_PRL_20, Sano_PRB_18, Chen_PRB_21, Shukla_PRB_18, Shukla_PRB_23}. In double perovskites having the general formula A$_2$$B$$B'$O$_6$, where A can be alkaline earth or rare-earth metals and $B$$B'$ are transition metals, this interplay is particularly rich because the degree of $B$-site ordering, octahedral distortions, and the nature of the magnetic exchange pathways can be modified through the choice of constituent cations \cite{Vasala_SSC_15, GarciaHernandez_PRL_01, Kumar_PRB1_20}. In rock-salt-ordered double perovskites containing only one magnetic $B$-site cation, the magnetic ions form a distinct sublattice and interact through extended $B$--O--$B'$--O--$B$ super-superexchange pathways mediated by the nonmagnetic $B'$ ions \cite{Vasala_SSC_15, deVries_PRL_10, Bos_PRB_04}. The competition among such exchange pathways can give rise to substantial magnetic frustration, making these compounds useful systems for exploring the connection between local structure and magnetic interactions \cite{Bos_PRB_04, Aharen_PRB_09, Nilsen_PRB_15}. More recently, a quantum spin liquid (QSL) state and Kitaev interactions have been anticipated in the paramagnetic (PM) phase of double perovskites that exhibit long-range magnetic ordering at low temperatures \cite{Singh_PRR2_20, Singh_PRB_21}. However, it is important to emphasize that an unquenched orbital magnetic moment is a prerequisite for having directional character in the different exchange interactions, owing to the non-spherical symmetry of various orbitals. Therefore, the 5$d$ transition metals, which exhibit strong spin-orbit coupling and therefore non-negligible orbital magnetic moments (particularly Ir-based compounds with $J_{\rm eff}=1/2$), have been extensively studied to understand the Kitaev model and, consequently, the QSL ground state with its exotic quasiparticle excitations \cite{Singh_PRR2_20, Singh_PRB_21, Glamazda_NatCom_16, Do_PRL_20, Kimchi_PRB_14, Cook_PRBR_15, Aczel_PRB_19}. 

In this context, it is important to note here that even cobalt based 3$d$ systems with comparatively weak spin-orbit coupling can retain an unquenched orbital magnetic moment if pairs of orbitals with similar symmetry are degenerate (at least within the statistical electronic fluctuations) \cite{Mabbs_book_73}. This is because any interconversion of electronic orbitals into one another by rotation about a given axis results in a non-zero orbital angular moment about that axis \cite{Mabbs_book_73}. The Co$^{2+}$ ions have a 3$d^7$ (t$_{2g}^5e_g^2$) configuration with $S=3/2$ and $L=3$, which gives a $^4F$ free-ion-like term and hence a triply degenerate  $^4T_1$ ground term in the presence of a weak crystal field limit, allowing a substantial unquenched orbital contribution to the magnetic moment \cite{Mabbs_book_73, Lloret_ICA_08}. The spin--orbit coupling further splits this manifold and produces a ground-state Kramers doublet that can be described by an effective pseudospin $\tilde{S}=1/2$, together with higher-lying multiplets. The coexistence of orbital degeneracy, spin--orbit coupling, and exchange anisotropy has motivated theoretical proposals of anisotropic, bond-dependent Kitaev type interactions in certain $d^7$ Co$^{2+}$ compounds \cite{Liu_PRB_18}. The realization of such interactions, however, depends sensitively on the local coordination and exchange geometry and cannot be inferred from the Co$^{2+}$ electronic configuration alone. 

Within the class of Co$^{2+}$ based 3$d$ materials, the SrLaCoNbO$_6$ having rock-salt-like Co/Nb ordering provides an interesting system for investigating the connection between local structure, lattice dynamics, and magnetism because Co$^{2+}$ is the only magnetic ion at $B$-site, whereas Nb$^{5+}$ is nonmagnetic \cite{Kumar_PRB1_20, Kumar_PRB2_20, Kumar_PRB_22}. The magnetization and specific heat measurements indicate antiferromagnetic (AFM) ordering of the Co$^{2+}$ moments below $T_{\rm N}=15$~K, with a significant contribution from the orbital magnetic moment in SrLaCoNbO$_6$, which  crystallizes in a nearly ordered monoclinic $P2_1/n$ structure \cite{Kumar_PRB1_20, Kumar_PRB2_20}. Further, neutron diffraction studies revealed ordering with a propagation vector $\mathbf{k}=(1/2, 0, 1/2)$ and significant magnetic frustration arising from competing 90$^\circ$ Co$^{2+}$--O--Nb--O--Co$^{2+}$ superexchange pathways \cite{Bos_PRB_04}. The relatively weak crystal-field splitting of high-spin Co$^{2+}$ preserves the essential $^4T_1$ character of the $3d^7$ configuration and associated Kramers doublet ground state with pseudo-spin $\tilde{S}=$ 1/2 \cite{Kumar_PRB2_20}. These characteristics make SrLaCoNbO$_6$ a suitable system for investigating how local structural modifications and magnetic correlations influence the lattice dynamics in the PM regime. 

Raman spectroscopy is particularly sensitive to such coupled degrees of freedom because it provides access to low-energy excitations, including phonons, magnons, and orbitons, and to their interactions through the temperature evolution of the mode frequencies, linewidth, intensities, and asymmetry in line shape \cite{Cottam_book_86, Ulrich_PRL_06, Ulrich_PRL_09, Ulrich_PRL_15, Kumar_PRB_12, Rovillain_PRB_10}. In magnetic materials, spin--phonon coupling can produce deviations of the phonon frequencies and linewidth from the conventional anharmonic behavior across the phase transitions \cite{Rovillain_PRB_10,  Laverdiere_PRB_06, Sohn_PRL_17}. Such deviations may also persist well above the long-range magnetic ordering temperature owing to short-range magnetic correlations, making Raman spectroscopy a sensitive probe to spin dynamics in the paramagnetic regime \cite{Singh_PRR1_20, Kamba_PRB_14, Iliev_APL_07}. Moreover, coupling of a discrete phonon to an underlying electronic, orbital, or magnetic continuum can additionally produce asymmetric Fano line shapes and a broad low-energy Raman response. Such features have been observed in a variety of correlated oxides and discussed in connection with mechanisms ranging from orbital and electronic fluctuations to magnetic correlations and, in selected spin-orbit-coupled systems, fractionalized magnetic excitations \cite{Ulrich_PRL_15, Singh_PRR2_20, Singh_PRB_21, Glamazda_NatCom_16, Sandilands_PRL_15, Nasu_NP_16, Chakkar_PRB_25}. Further, the extended x-ray absorption fine structure (EXAFS) provides a local, element-specific probe of the coordination environment where the temperature-dependent changes in the  bond distances and Debye--Waller factors can reveal subtle local structural modifications accompanying anomalous lattice dynamics \cite{Rodrigues_JMCC_22, Bridges_PRB_07, Panchwanee_PCM_19}.  

Therefore, in this work, we investigate the lattice dynamics and local structural evolution of the ordered double perovskite SrLaCoNbO$_6$ using temperature-dependent Raman spectroscopy and Co $K$-edge EXAFS measurements. Interestingly, we observe clear deviations from conventional anharmonic behavior in several optical phonon modes, with pronounced anomalies well above the long-range AFM ordering temperature ($T_{\rm N} \approx 15$~K). In addition, selected low-energy modes exhibit distinct Fano line shapes accompanied by a broad low-energy continuum, pointing to a coupling between discrete phonons and an underlying fluctuation continuum. Moreover, the EXAFS measurements reveal correlated changes in the local Co--O bond distances and Debye--Waller factors, indicating subtle modifications in the local CoO$_6$ environment. These results reveal a close correlation between anomalous lattice dynamics and local structural evolution in SrLaCoNbO$_6$. We discuss several possible microscopic origins of these observations, including local structural distortions as well as the magnetic, orbital, and localization related correlations within the broader physical framework.  

\section{\noindent ~Experimental}

A polycrystalline SrLaCoNbO$_6$ sample was synthesized by the conventional solid-state route, the details of which are provided in \cite{Kumar_PRB1_20}. The Rietveld refinement of the room-temperature x-ray diffraction pattern shows that the sample crystallizes in a nearly B-site ordered monoclinic structure with space group $P2_1/n$, consistent with Refs.~\cite{Bos_PRB_04, Kumar_PRB1_20} (also see Figs.~S1(a, b) of \cite{SI}). 

Temperature-dependent Raman spectroscopy was performed in the temperature range 12–300~K using the 633~nm line of a He–Ne laser and the 514~nm line of an Ar-ion laser. The spectra were recorded using a Dilor XY triple-monochromator spectrometer in backscattering geometry and detected with a liquid nitrogen cooled CCD camera. A laser power of 5~mW with a spot size of $\sim$0.1~mm was used to avoid sample heating. The Raman spectra at each temperature were calibrated using standard neon lines to accurately determine the positions of the various modes and the instrumental resolution. 

Low-temperature x-ray absorption spectroscopy (XAS) measurements were carried out in the temperature range 21–300~K at the Co $K$ edge in transmission mode at the BL-09 (4–25~keV) beamline of the Indus-2 synchrotron source (2.5~GeV, 200~mA) at the Raja Ramanna Centre for Advanced Technology (RRCAT), Indore, India. At each temperature, three spectra were collected, averaged, and used for analysis. Each spectrum was calibrated using a standard Co metal foil, following the procedure outlined in Ref.~\cite{Kumar_PRB_22}.

\section{\noindent ~Results }

\subsection{\noindent ~Low temperature Raman spectroscopy:}

\begin{figure}
\centering
\includegraphics[width=1.03\linewidth]{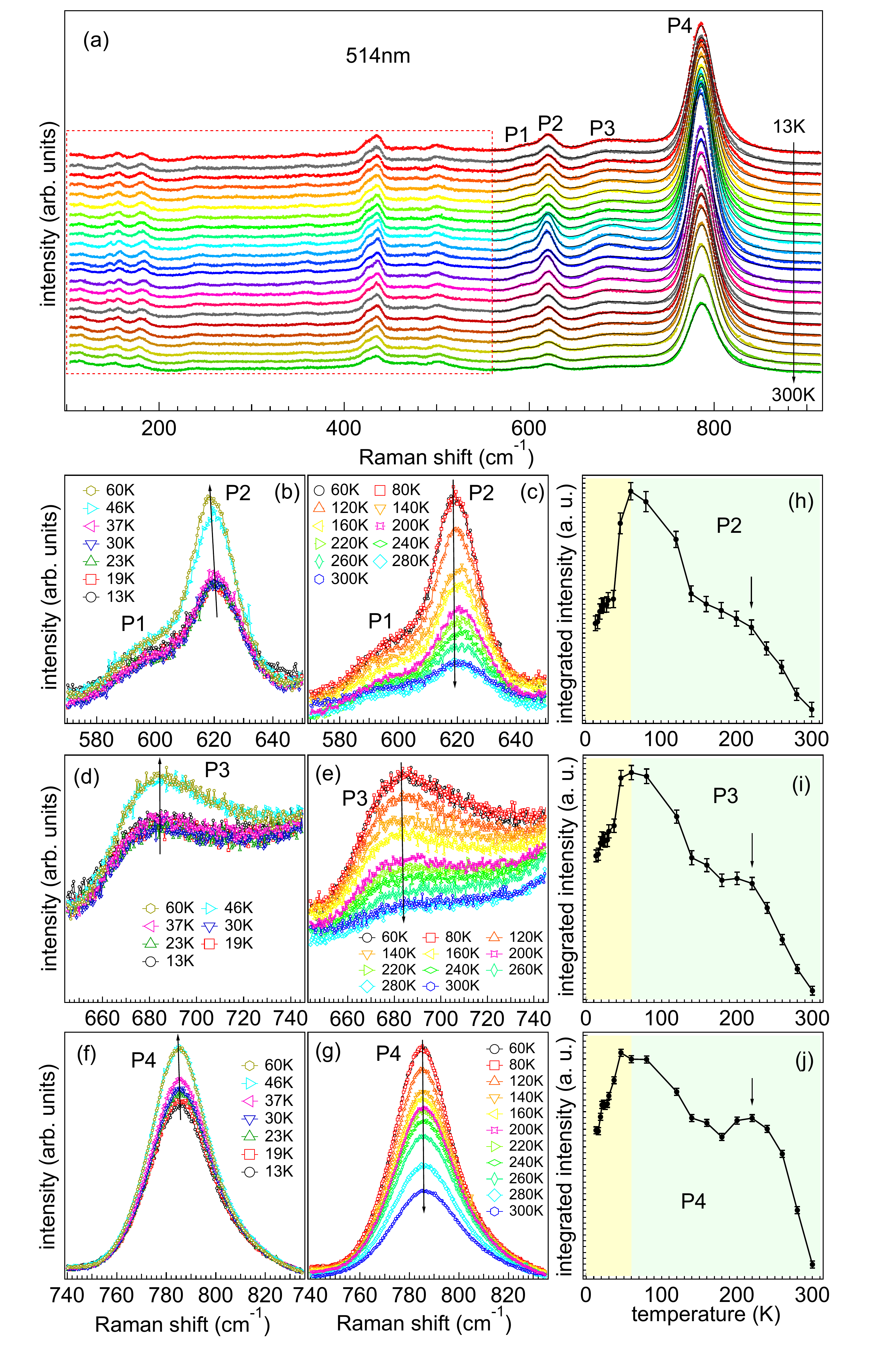}
\caption {(a) The Raman spectra of SrLaCoNbO$_6$ measured between 13 and 300~K using the 514~nm line from an Ar-ion laser. Each spectrum is vertically offset for clarity. The solid black curves represent fits to the spectra in the 550–850~cm$^{-1}$ range using four Voigt functions corresponding to modes P1--P4. (b, c) Temperature evolution of the P1 and P2 modes, (d, e) P3 mode, and (f, g) P4 mode at selected temperatures, where the arrows indicate the direction of increasing temperature. (h–j) Temperature dependent integrated intensities of modes P2, P3, and P4, respectively. } 
\label{2_514_zoom2}
\end{figure}

Figure~\ref{2_514_zoom2}(a) shows the Raman spectra of SrLaCoNbO$_6$ measured between 13 and 300~K using a 514~nm excitation. The four modes observed in the 550–850~cm$^{-1}$ range (P1–P4) are assigned to internal stretching vibrations of the (Co/Nb)–O bonds within the corner-sharing (Co/Nb)O$_6$ octahedra. Raman modes in the 200–550~cm$^{-1}$ region originate from bending vibrations of the (Co/Nb)O$_6$ octahedra, while those below 200~cm$^{-1}$ are mainly associated with translational motions of the A-site ions \cite{Andrews_DT_15, Iliev_PRB_07, Ayala_JAP_07}. SrLaCoNbO$_6$ crystallizes in a B-site ordered monoclinic structure with the space group $P2_1/n$ and the point symmetry C$_{2h}$, exhibiting a b$^-$b$^-$c$^+$ tilt pattern of the octahedra \cite{Bos_PRB_04}. The group theory predicts a total of 24 Raman-active (12 A$_g$ and 12 B$_g$) and 36 infrared-active modes for this symmetry, as displayed in Table~S1 of \cite{SI}, in the irreducible representation \cite{Singh_PRR1_20, Singh_PRM_20}. The 24 Raman-active modes can be further classified as $\Gamma_{\rm Raman}$ = $\nu_1$(A$_g$+B$_g$) + $\nu_2$(2A$_g$+2B$_g$) + $\nu_5$(3A$_g$+3B$_g$) + T(3A$_g$+3B$_g$) + L(3A$_g$+3B$_g$) \cite{Andrews_DT_15, Ajay_JAP20}. Here, $\nu_1$, $\nu_2$, and $\nu_5$ are the internal stretching modes within the (Co/Nb)O$_6$ octahedra, T represents the translational modes due to the motion of A-site (La or Sr) cations, and L represents the librational modes corresponding to the octahedral tilting. The $\nu_1$ and $\nu_2$ modes represent the symmetric and asymmetric stretching of the (Nb/Co)-O bonds, respectively, and $\nu_5$ represents the oxygen bending motion in the octahedra, i.e., the O--Nb/Co--O vibrations.

\begin{figure}[h] 
\centering
\includegraphics[width=1.05\linewidth]{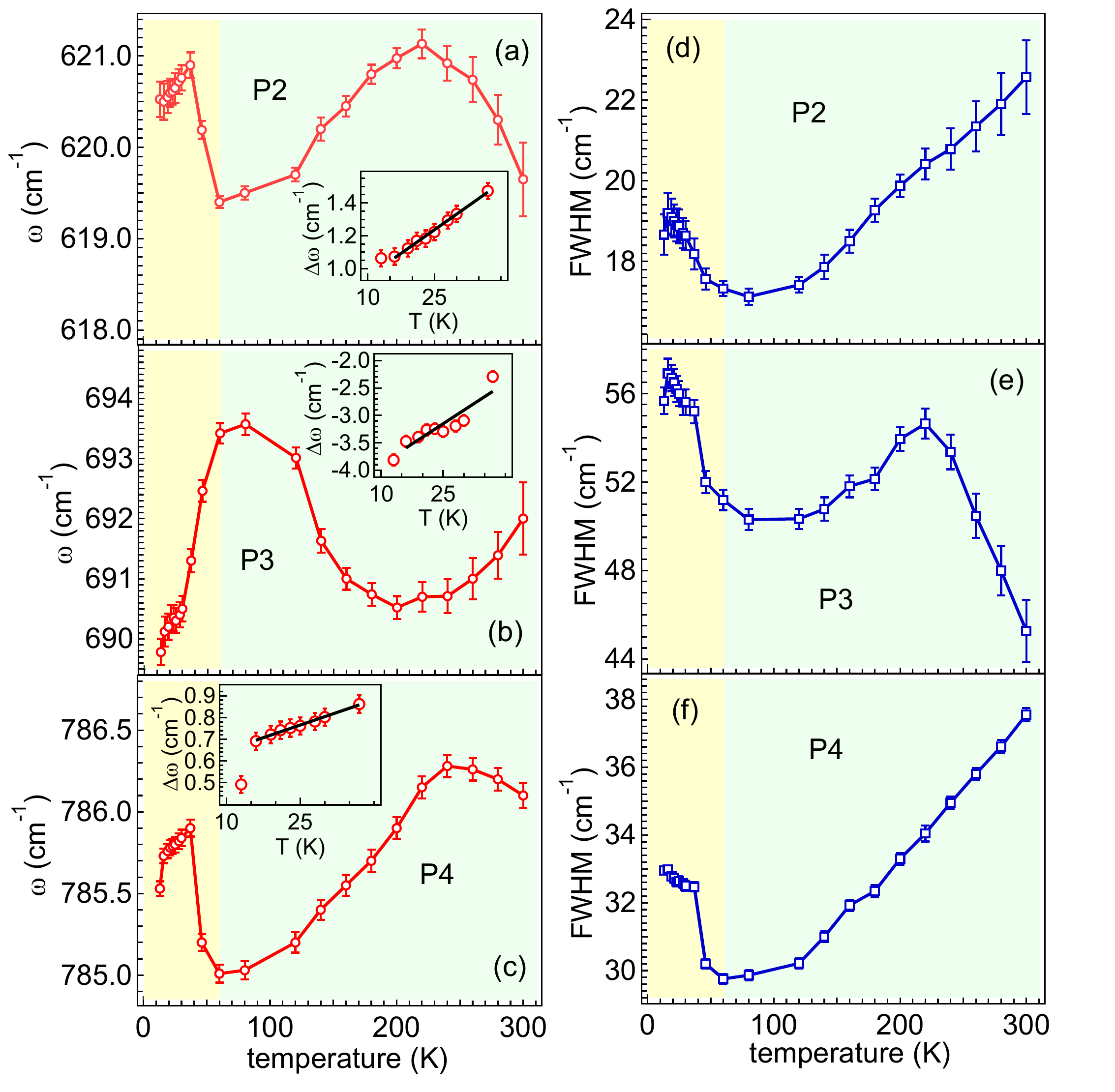}
\caption {Temperature dependence of the (a--c) phonon frequencies and (d--f) linewidths of modes P2--P4. The insets in (a--c) display the deviation of the mode frequency ($\Delta\omega$) below $\sim 60$~K. The solid black lines are guides to the eye.} 
\label{3_514_fit2}
\end{figure}

\begin{figure*} 
\centering 
\includegraphics[width=7in]{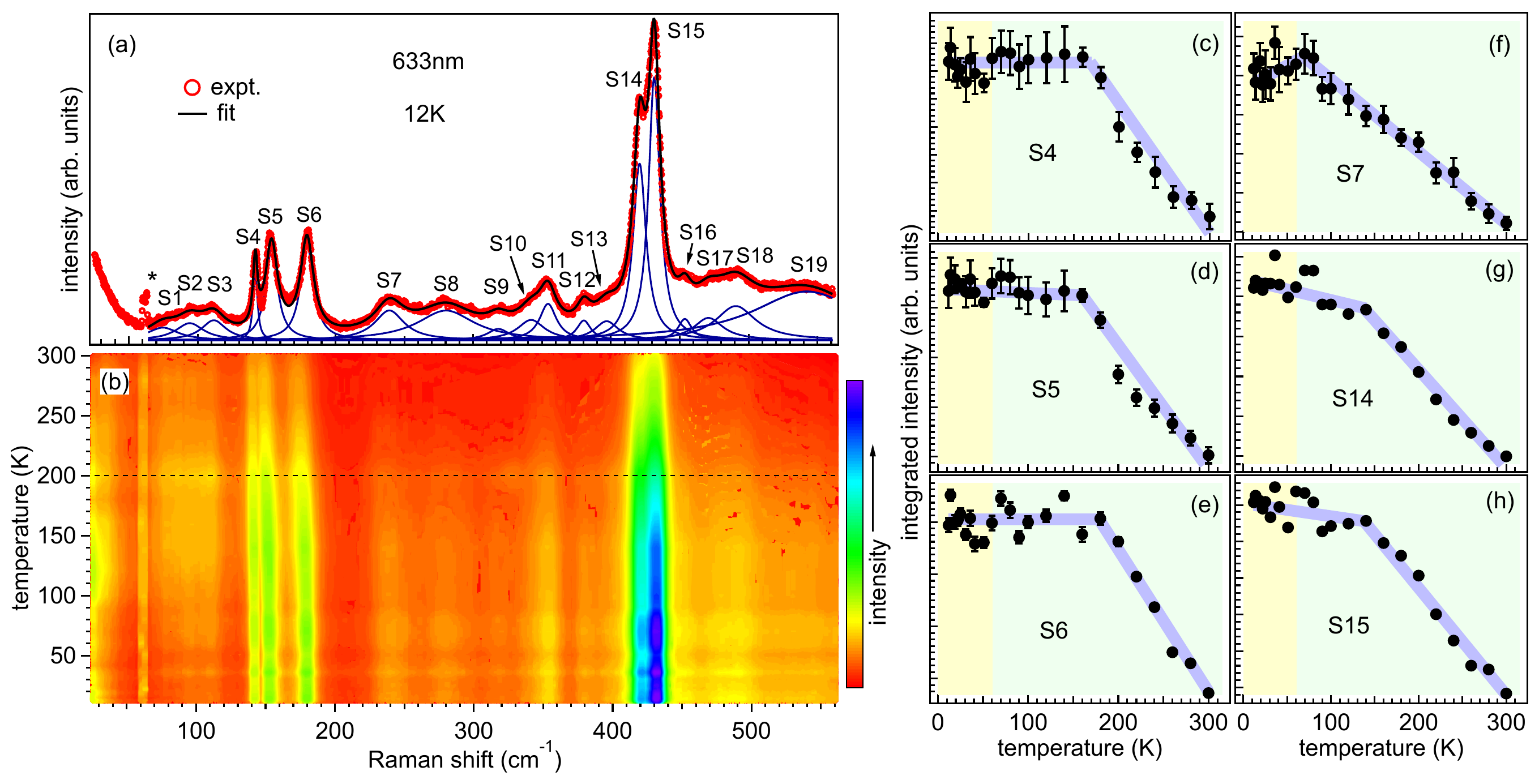}
\caption {Raman spectrum of SrLaCoNbO$_6$ measured at 12 K using the 633 nm laser line as  excitation source. The solid black and blue lines represent the cumulative fit and the individual Voigt components corresponding to the different modes, respectively. A sharp feature near 60 cm$^{-1}$, marked with an asterisk, originates from the Ne emission line. (b) Contour plot of the Raman spectra over the temperature range 12--300~K. The horizontal dashed line indicates a change in the intensity of the low-energy modes around 200~K. (c--h) Temperature dependence of the integrated intensities of the prominent modes. The blue lines are guides to the eye.}
\label{4_633_LT}
\end{figure*}

A significant overlap among various bending and translational modes is observed below $\sim$550~cm$^{-1}$ [highlighted by the red dashed rectangle in Fig.~\ref{2_514_zoom2}(a)], which complicates a reliable analysis of these modes (see Fig.~S2 of Ref.~\cite{SI}). We therefore first focus on the temperature evolution of the high-energy modes P1–P4. Interestingly, the intensity of all these modes increases with temperature up to $\sim$60~K and decreases upon further heating, as shown in Figs.~\ref{2_514_zoom2}(b–g). This behavior suggests local structural modifications or interaction of lattice vibrations  (phonons) with other degrees of freedom across this temperature range. To quantify this effect, each spectrum in the 550–850~cm$^{-1}$ range is fitted using a sum of four Voigt functions and a linear background, as illustrated by the solid black curves in Fig.~\ref{2_514_zoom2}(a). The resulting integrated intensities of modes P2--P4 are shown in Figs.~\ref{2_514_zoom2}(h–j), respectively, revealing an enhancement up to around 60~K followed by a gradual decrease at higher temperatures. In addition, a second anomaly is observed around 200--220~K, which becomes more pronounced with increasing mode frequency, as indicated by the arrows in Figs.~\ref{2_514_zoom2}(h--j). The extracted frequencies ($\omega$) and FWHM ($\Gamma$) of modes P2–P4 are presented in Figs.~\ref{3_514_fit2}(a–c) and \ref{3_514_fit2}(d–f), respectively. All these modes exhibit an abrupt renormalization of both $\omega$ and $\Gamma$ around 60~K, indicating a change in the phonon self-energy below this temperature. Modes P2 and P4 show a sharp, step-like jump in peak position near 60~K, followed by a red shift upon further cooling, whereas mode P3 exhibits a large and continuous red shift below 60~K [see Figs.~\ref{3_514_fit2}(a–c)]. The deviation of the mode frequencies from the high-temperature behavior (see Figs.~S3(a–c) of Ref.~\cite{SI}) is shown in the insets of Figs.~\ref{3_514_fit2}(a–c), where the solid black lines serve as guides to the eye. The sudden change in mode frequency below 60~K is accompanied by an abrupt linewidth broadening for all three modes, as shown in Figs.~\ref{3_514_fit2}(d–f). Since the linewidth ($\Gamma$) of a Raman mode is inversely proportional to the phonon lifetime ($\tau$), the observed increase in $\Gamma$ below 60~K is consistent with coupling of the optical phonons to an additional degree of freedom, which may be lattice, electronic, or magnetic in origin. 

\begin{figure*}
\includegraphics[width=\linewidth]{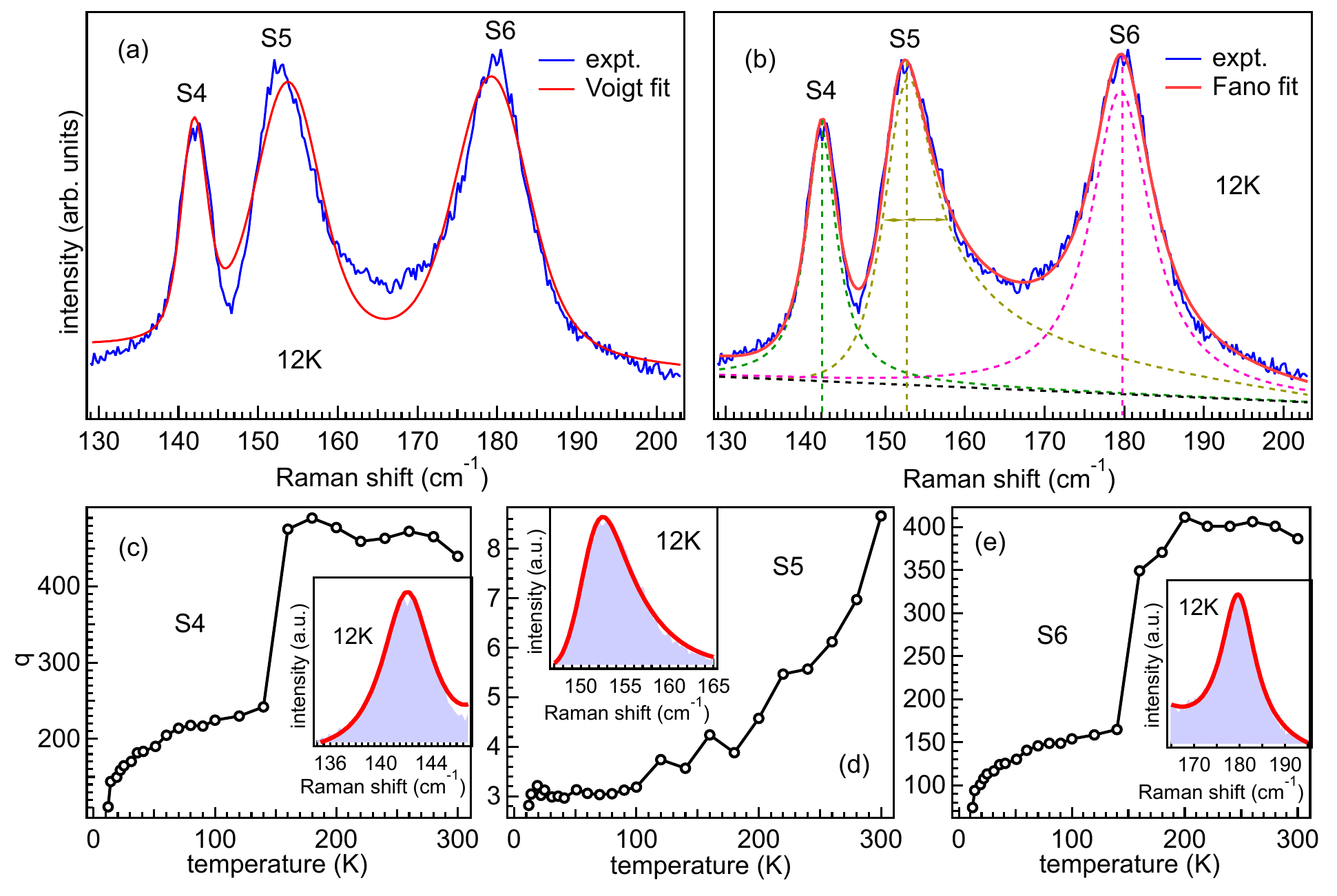}
\caption {Fits to the S4-S6 modes of the Raman spectrum at 12~K using the sum of (a) Voigt (b) Fano line shapes. (c--e) Temperature dependence of the Fano asymmetry parameter, $q$ for the S4--S6 modes, respectively. The insets show individual Fano fits to the experimental data at 12~K.} 
\label{5_633_LT_zoom5}
\end{figure*}

Notably, mode P3 exhibits substantially larger linewidth than P2 and P4 across the entire temperature range, including at low temperatures where anharmonic phonon–phonon scattering is strongly suppressed [see Figs.~\ref{3_514_fit2}(d–f)]. This behavior suggests a dominant contribution from lattice disorder rather than purely intrinsic anharmonicity. Random A-site occupancy (Sr/La) and/or local distortions of the (Co/Nb)O$_6$ octahedra likely introduce spatial fluctuations in bond lengths and force constants, resulting in inhomogeneous broadening via a distribution of local vibrational frequencies. The comparatively narrower linewidths of P2 and P4 indicate their weaker sensitivity to such structural disorder, consistent with distinct vibrational characters of different P modes. Furthermore, modes P2 and P4 display an anomalous red shift, whereas mode P3 exhibits a blue shift above 200–220~K [see Figs.~\ref{3_514_fit2}(a–c)]. The linewidth also shows a distinct change in this temperature range for mode P3, while no notable deviation is observed for modes P2 and P4 [see Figs.~\ref{3_514_fit2}(d–f)]. This additional renormalization of the phonon self-energy parameters, together with the corresponding changes in integrated intensity discussed above, points to an another interaction of phonons with the spin system and/or lattice degrees of freedom in this temperature range.

To understand the temperature evolution of the low-frequency phonon modes, high-resolution Raman spectra were collected in the range of 25--560~cm$^{-1}$ using the 633~nm line of a He--Ne laser [see Figs.~\ref{4_633_LT}(a, b)]. The low-frequency modes S1--S6 arise from translational motions of the A-site (Sr and/or La) ions, whereas S7--S18 correspond to O--Co/Nb--O bending vibrations along the adjacent (Co/Nb)O$_6$ octahedra. The intensities of the translational modes (S1--S6) remain nearly constant up to $\sim$180--200~K [horizontal dashed line in Fig.~\ref{4_633_LT}(b)], followed by a pronounced decrease toward 300~K. The temperature dependence of the integrated intensity of selected modes (S4--S7, S14 and S15) is shown in Figs.~\ref{4_633_LT}(c--h). A pronounced reduction in the intensity of modes S4--S6 is observed around 180--200 K, which may be associated with local lattice distortions and/or coupling to spin or electronic degrees of freedom in this temperature range. In contrast, the high-energy bending modes S14 and S15 exhibit a more gradual intensity reduction beginning near $\sim$150~K. Notably, mode S7 shows a distinct deviation around 60~K, indicating the notable influence of the low temperature coupling on the bending modes as well. This anomaly is, however, weaker for the low-frequency S modes than for the stretching modes P1--P4 discussed above. Moreover, an upturn is observed below $\sim$60~cm$^{-1}$ [see Figs.~\ref{4_633_LT}(a)], indicating the presence of a quasielastic scattering component in the spectra, discussed later. 

\begin{figure*}  
\centering
\includegraphics[width=\linewidth]{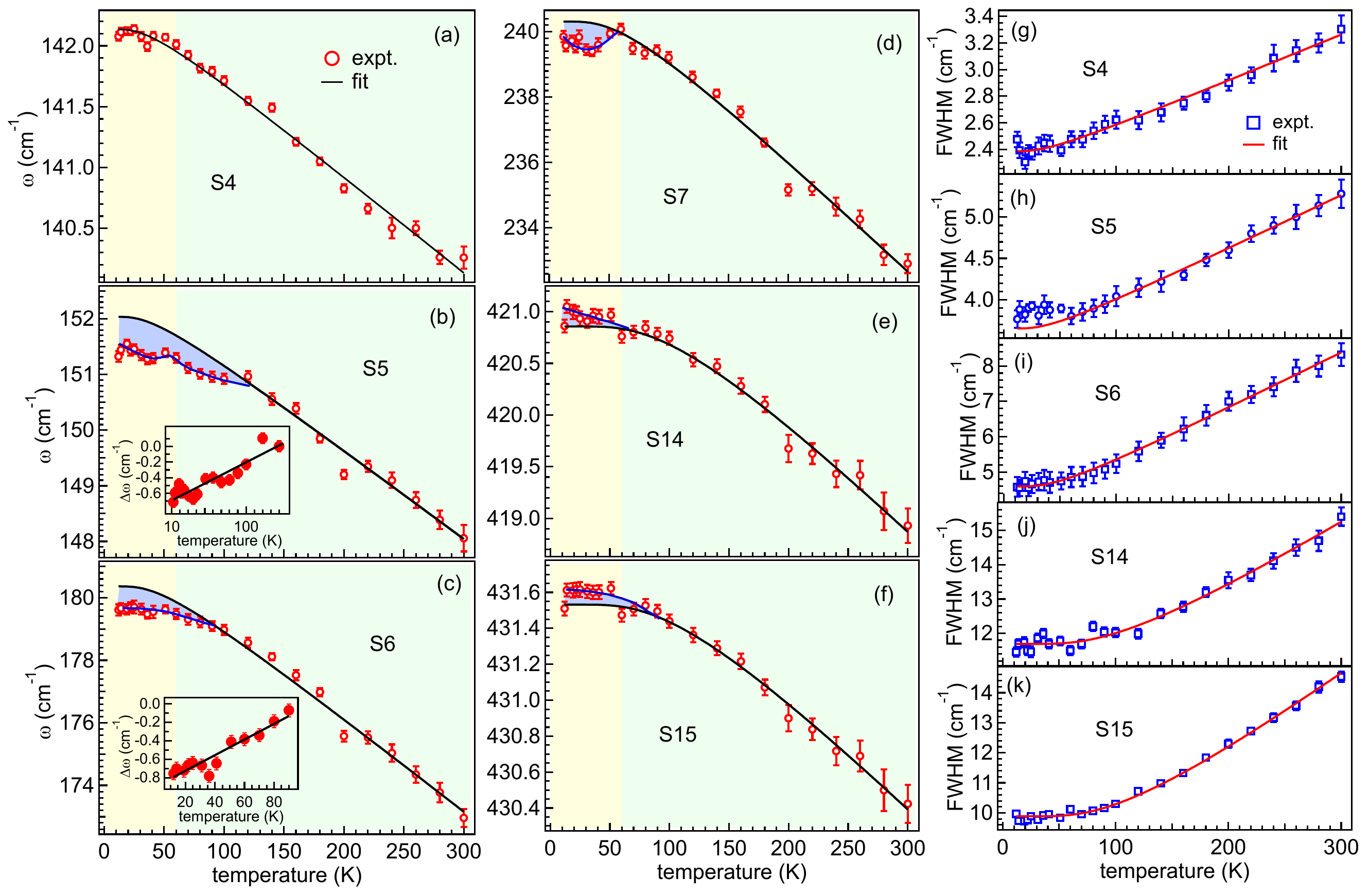}
\caption {(a--f) Temperature dependence of the phonon frequencies of modes S4--S7, S14, and S15, respectively. The solid black curves represent fits to the experimental data using the three-phonon anharmonic decay model [Eq.~(2)], and the shaded blue regions highlight deviations from this model. The insets in (b) and (c) show the deviation from the anharmonic model ($\Delta\omega$) in the low-temperature range; the black lines serve as guides to the eye. (g--k) Temperature dependence of the linewidth (FWHM) of these modes. The solid red curves represent fits using the anharmonic phonon decay model.} 
\label{7_633_LT_fit}
\end{figure*} 

The line shape of low-energy phonon modes provides key insight into their coupling to other degrees of freedom  \cite{Sandilands_PRL_15, Glamazda_NatCom_16, Singh_PRR2_20, Glamazda_PRB_17}. Interestingly, all three prominent low-energy phonon modes, S4--S6, exhibit clear asymmetry that cannot be adequately captured by a sum of Voigt functions, as illustrated in Fig.~\ref{5_633_LT_zoom5}(a) at $T = 12$~K. Instead, the spectra are well described by Fano profiles, reflecting coupling of the discrete phonons to an underlying electronic or spin continuum \cite{Sandilands_PRL_15, Glamazda_NatCom_16}. The Fano line shape is given by \cite{Fano_PR_124}
\begin{equation}
I(\omega) = I_0 \frac{(q + \epsilon)^2}{1 + \epsilon^2},
\end{equation}
where $\epsilon = (\omega - \omega_0)/\Gamma$, $\omega_0$ is the bare phonon frequency, $\Gamma$ the linewidth, and $q$ the asymmetry parameter. A sum of three Fano functions with a linear background  accurately reproduces the S4--S6 modes, as shown in Fig.~\ref{5_633_LT_zoom5}(b). Mode S5 exhibits the strongest asymmetry, while S4 and S6 display relatively weaker but non-negligible asymmetries. This difference reflects the distinct intrinsic origins of these modes and, consequently, their varying degrees of coupling to the  continuum. The inverse Fano asymmetry parameter, $1/q$, quantifies the strength of this coupling: a highly asymmetric peak ($q \to 0$) implies strong coupling, whereas in the weak-coupling limit ($q \to \infty$), the Fano profile approaches a symmetric Lorentzian line shape. Figs.~\ref{5_633_LT_zoom5}(c--e) show the temperature dependence of the Fano asymmetry parameter $q$ for modes S4--S6, respectively (see Fig.~S4 of \cite{SI} for temperature-dependent Fano fits). Notably, modes S4 and S6 exhibit $q$ values approximately two orders of magnitude larger than that of mode S5, indicating significantly weaker coupling to the underlying continuum. The insets of Figs.~\ref{5_633_LT_zoom5}(c--e) present representative Fano fits to the individual phonon modes at 12~K. Interestingly, an abrupt jump in the Fano asymmetry parameter for modes S4 and S6 is observed around 160~K, corresponding to an approximately 100\% increase in $q$. Mode S5 also shows a more rapid increase in $q$ above this temperature, suggesting a change in the nature of the phonon coupling to spin/electronic excitations above $\sim$160~K.

The high-energy modes S7--S19 are symmetric, and therefore were fitted using Voigt profiles, which take the instrument resolution also into account (see Fig.~S4 in Ref.~\cite{SI}). The peak positions and FWHM of the prominent modes are presented in Figs.~\ref{7_633_LT_fit}(a--f) and \ref{7_633_LT_fit}(g--k), respectively. The temperature dependence of the phonon self-energy parameters is analyzed using an anharmonic phonon–phonon interaction model based on a three-phonon process, i.e., the symmetric decay of an optical phonon into two lower-energy phonons, given as~\cite{Balkanski_PRB_83, Shukla_PRB22} 
\begin{equation}
\omega_{\rm anh}(T) = \omega_0 + A\left(1 + \frac{2}{exp(\hbar \omega_0/2k_B T) - 1} \right),
\end{equation}
\begin{equation}
\Gamma_{\rm anh}(T) = \Gamma_0 + B\left(1 + \frac{2}{exp(\hbar \omega_0/2k_B T) - 1} \right),
\end{equation}
where \( k_B \), \( \omega_0 \), and \( \Gamma_0 \) are the Boltzmann constant, mode frequency, and linewidth at 0~K, respectively. \( A \) and \( B \) are the constants related to the three-phonon anharmonic decay channels. The solid lines in Figs.~\ref{7_633_LT_fit}(a--k) represent the best fit curves obtained using the above models, and the corresponding parameters are listed in Table~S2 of Ref.~\cite{SI}. We observe that the anharmonic three-phonon decay model satisfactorily describes the temperature dependence of the mode frequencies for $T \gtrsim 100$~K [see Figs.~\ref{7_633_LT_fit}(a--f)]. However, the translational modes S5--S7 exhibit an additional redshift [see Figs.~\ref{7_633_LT_fit}(b--d)], while the bending modes S14 and S15 display a weak blueshift at low temperatures [see Figs.~\ref{7_633_LT_fit}(e, f)], highlighted by the blue shaded regions. Modes S7, S14, and S15 deviate from the anharmonic model around 60~K, analogous to the behavior observed for the high-energy P1--P4 modes discussed above. In contrast, modes S5 and S6 show deviations even above 60~K, suggesting their entanglement with an additional electronic or spin degree of freedom, as reflected in their Fano line shape. The deviations in the peak positions of these modes are shown in the insets of Figs.~\ref{7_633_LT_fit}(b, c), where the black solid lines serve as guides to the eye. No significant deviations from the anharmonic three-phonon decay model are observed in the FWHM down to 12~K [Figs.~\ref{7_633_LT_fit}(g--k)], except for mode S5, which exhibits a very weak linewidth broadening below $\sim$60~K [see Fig.~\ref{7_633_LT_fit}(h)].

\begin{figure} 
\centering
\includegraphics[width=3.35in]{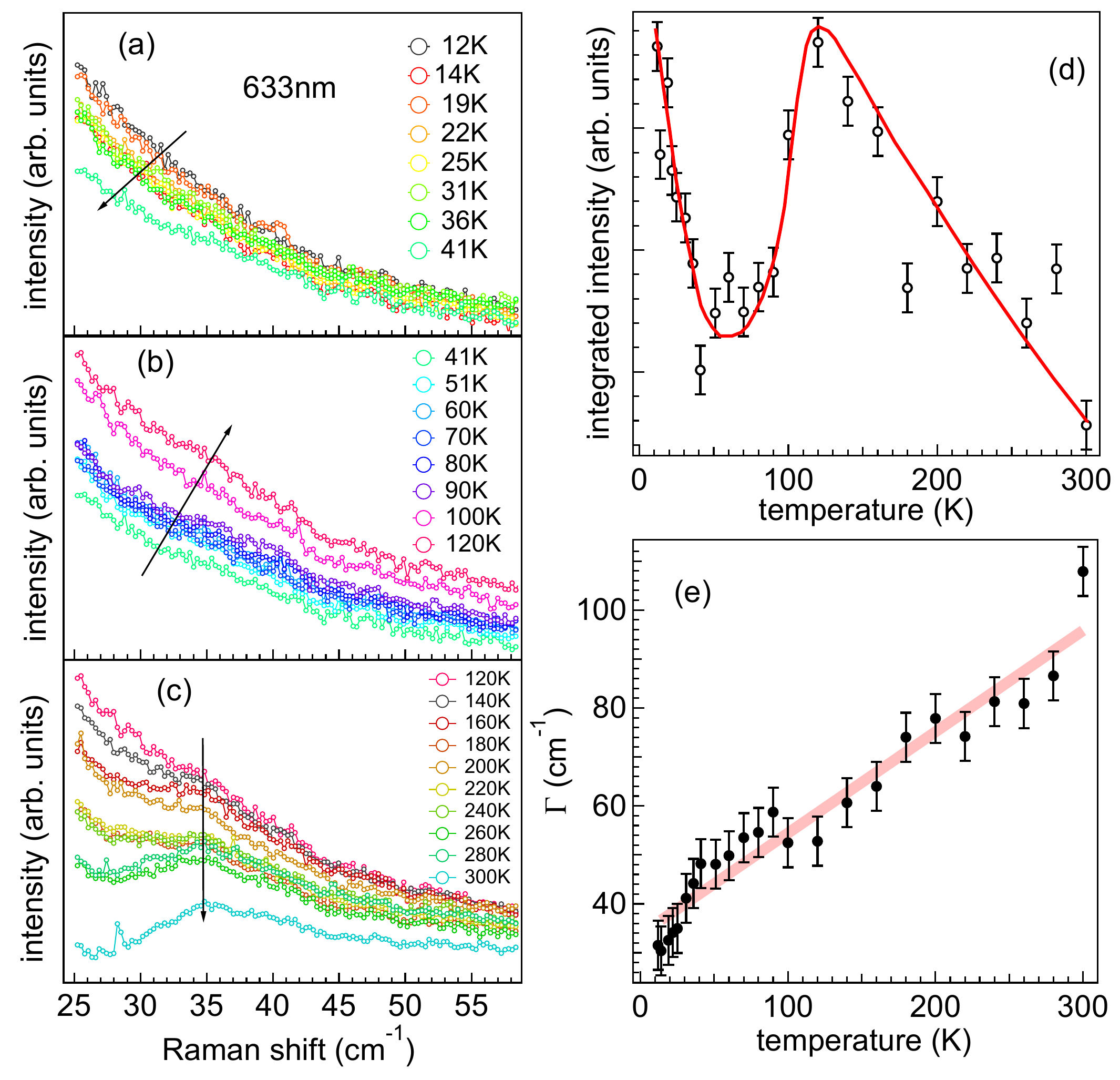}
\caption {(a--c) Raman spectra of SrLaCoNbO$_6$ in the range 25 to 60~cm$^{-1}$ at different temperatures; the arrows indicate the direction of increasing temperature. (d) Temperature dependence of the integrated intensity of the quasielastic peak; the solid red curve serves as a guide to the eye. (e) Temperature dependence of the FWHM of the low-energy quasielastic response; the red line serves as a guide to the eye. } 
\label{5_633_LT_zoom3}
\end{figure}

We now focus on the low-energy quasielastic response (central peak) observed in SrLaCoNbO$_6$ and its temperature dependence, as shown in Figs.~\ref{5_633_LT_zoom3}(a–c). The quasielastic tail, appearing below $\sim$60~cm$^{-1}$, initially decreases in intensity up to $\sim$40~K, then increases to a maximum around 120~K, and subsequently decreases upon further heating. The low-energy quasielastic tail in Raman spectra originates from  fluctuation processes that scatter light without creating or annihilating well-defined phonons. Its most common source is relaxational dynamics—fluctuations on time scales longer than the phonon periods—arising from structural or orientational disorder \cite{Milkus_PRB_16, Menahem_PRM_23, Michel_PRB_78, Sokoloff_PRB_90, Halperin_PRB_76}. In addition, spin fluctuations in magnetic systems \cite{Choi_PRB_11, Wulferding_PRB_10, Yamada_PRB_94} and charge-density fluctuations in conductors or correlated electron systems can contribute to the broad central peak in Raman spectra. Since SrLaCoNbO$_6$ is highly insulating \cite{Kumar_PRB1_20}, charge-density fluctuations should be small and therefore can be excluded as a possible origin. Therefore, this feature is primarily attributed to octahedral tilting and dynamic lattice fluctuations associated with local metastability, possibly induced by A-site disorder arising from the small ionic-size mismatch between twelvefold-coordinated La$^{3+}$ and Sr$^{2+}$ ions \cite{Shannon_AC_79}. Moreover, isotopic disorder at the Sr site ($^{84}$Sr–$^{88}$Sr), and to a lesser extent oxygen isotope disorder, may introduce local mass fluctuations and subtle symmetry breaking, further enhancing the central peak. In the present case, dynamic short-range spin correlations in the paramagnetic state, originating from frustrated low-temperature AFM interactions in SrLaCoNbO$_6$ \cite{Bos_PRB_04}, may also contribute to this low-energy Raman response \cite{Choi_PRB_11, Wulferding_PRB_10, Yamada_PRB_94}.

To analyze the temperature evolution of the quasielastic Raman response (central peak), Lorentzian profiles are fitted to the spectra \cite{Choi_PRB_11, Wulferding_PRB_10, Yamada_PRB_94, Kuroe_PRB_97} centered at $\omega = 0$ (see Fig.~S5 of Ref.~\cite{SI}). The extracted temperature dependences of the integrated intensity and FWHM, evaluated over the range $-60$ to $60~\mathrm{cm}^{-1}$, are shown in Figs.~\ref{5_633_LT_zoom3}(d) and \ref{5_633_LT_zoom3}(e), respectively. The integrated intensity decreases upon warming from $12~\mathrm{K}$ to $\sim 60~\mathrm{K}$, increases up to $\sim 120~\mathrm{K}$, and then gradually decreases toward $300~\mathrm{K}$. In contrast, the FWHM increases nearly linearly with temperature, as indicated by the solid red line in Fig.~\ref{5_633_LT_zoom3}(e). Interestingly, an additional asymmetric feature emerges near $\sim 35~\mathrm{cm}^{-1}$ for $T\gtrsim$160~K, see Fig.~\ref{5_633_LT_zoom3}(c), with its intensity increasing continuously up to $300~\mathrm{K}$. We attribute this feature to a temperature-induced subtle modification of the $(\mathrm{Co/Nb})\mathrm{O}_{6}$ octahedra. This low-energy peak was excluded from the Lorentzian fit to ensure a reliable comparison of the central-peak parameters. The onset of this structural modification in SrLaCoNbO$_6$ near this temperature is consistent with the pronounced variations in the intensities of both the low-energy (S) and high-energy (P) phonon modes discussed above and is further supported by the temperature-dependent EXAFS measurements presented below. 

\begin{figure*}
\centering
\includegraphics[width=\linewidth]{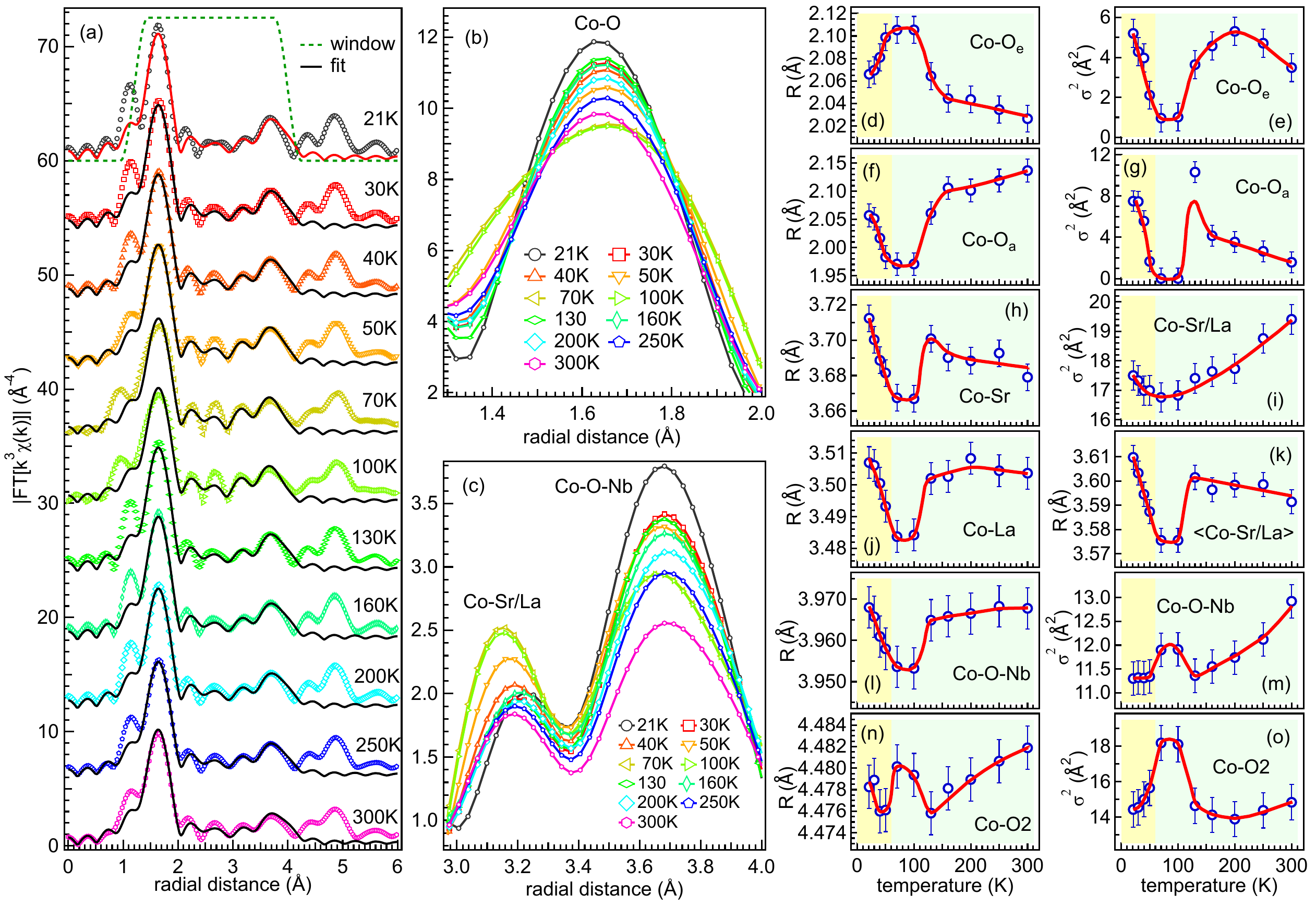}
\caption {(a) Temperature-dependent magnitude of the Fourier-transformed $k^3\chi(k)$ data of the Co K-edge EXAFS spectra of SrLaCoNbO$_6$. Each spectrum is cumulatively shifted vertically upward by 6~\AA$^{-4}$ relative to that recorded at 300~K for clarity of presentation. The solid black and dashed green lines represent the best fit to the experimental data and the fitting window, respectively.  (b, c) The enlarged view of the $\chi$(R) data in two different spectral regions. (d--o) Temperature dependence of the different scattering path lengths and their respective Debye-Waller factors (multiplied by 10$^3$), extracted from the EXAFS fitting of SrLaCoNbO$_6$ sample. The solid red lines serve as guides to the eye.} 
\label{11_EXAFS_fit}
\end{figure*} 

\subsection{\noindent ~Low temperature X-ray absorption studies: }

To investigate possible subtle changes in the local coordination environment around Co at low temperatures, temperature-dependent extended x-ray absorption fine structure (EXAFS) measurements were performed at the Co K edge for SrLaCoNbO$_6$ (see Figs.~S6(a, b) of Ref.~\cite{SI}). Figure~\ref{11_EXAFS_fit}(a) shows the amplitude of the Fourier-transformed $k^3\chi(k)$ spectra at selected temperatures. A Hanning window was applied over the $k$ range 3.7–12~\AA$^{-1}$, with a background cutoff $R{\text{bkg}} = 1.2$~\AA, to obtain the corresponding $\chi(R)$ spectra. The most intense peak, located near 1.5~\AA\ (phase uncorrected), arises from Co–O single-scattering paths within the CoO$_6$ octahedra. In the monoclinic $P2_1/n$ structure, the six Co–O bonds split into three pairs of inequivalent bond lengths \cite{Bos_PRB_04}. However, the differences among the four equatorial Co–O bonds in the $xy$ plane are smaller than the real-space resolution of the present EXAFS data \cite{Bos_PRB_04, Kumar_JPCL_22}. Therefore, the four Co--O$_e$ bonds were treated as equal, while the two axial Co--O$_a$ bonds along the $z$ axis were fitted separately to describe the distorted CoO$_6$ octahedra. The weak feature near 3.2~\AA\ is attributed to Co–Sr and Co–La single-scattering paths from the $A$-site ions located at the center of the pseudocubic perovskite unit cell. Given their comparable bond lengths and scattering mechanism, a common Debye–Waller factor ($\sigma^2$) was used for both Co–Sr and Co–La paths. The peak around 3.7~\AA\ corresponds to Co–Co/Nb scattering. Since the antisite disorder between Co and Nb in SrLaCoNbO$_6$ is reported to be less than 10\% \cite{Bos_PRB_04, Kumar_PRB1_20, Kumar_PRB_22}, a fully $B$-site-ordered configuration was adopted in the fitting, with six Co–Nb scattering paths. In the perovskite lattice, each Co–Nb path is mediated by an intervening oxygen atom \cite{Ajay_PRB24}. Accordingly, instead of using a direct Co–Nb single-scattering path, the Co–O–Nb forward multiple-scattering contribution was included in the fit. The feature near 3.7~\AA\ exhibits a clear asymmetry toward higher radial distances [see Fig.~\ref{11_EXAFS_fit}(a)], which is attributed to additional Co–O2 scattering from second-nearest-neighbor oxygen atoms and was therefore incorporated into the fitting model. The $\chi(R)$ spectrum was first fitted at the lowest measured temperature (21~K). The amplitude reduction factor ($S_0^2$) and the energy shift ($E_0$) determined from this fit were subsequently fixed, and only the path lengths and the corresponding Debye–Waller factors were allowed to vary at all other temperatures.

The best fit of the model function to the $\chi(R)$ data in the range 1.2 to 4~\AA\ [indicated by the dotted green window in Fig.~\ref{11_EXAFS_fit}(a)] at different temperatures is shown by the solid curves in Fig.~\ref{11_EXAFS_fit}(a). The temperature evolution of the fitting parameters for the different scattering paths is plotted in Figs.~\ref{11_EXAFS_fit}(d--o). Interestingly, the strength of the peak corresponding to the Co--O path in the $\chi(R)$ spectra decreases with increasing temperature up to $\approx 60~\mathrm{K}$, then increases abruptly between 100 and 120~K, and subsequently decreases again for $T \geqslant 160~\mathrm{K}$, as shown in Fig.~\ref{11_EXAFS_fit}(b) (see also Figs.~S7(a--c) and S7(g) in Ref.~\cite{SI} for clarity). The temperature-induced changes in the Co--O$_e$ and Co--O$_a$ bond distances and their corresponding Debye--Waller (DW) factors are shown in Figs.~\ref{11_EXAFS_fit}(d--g). The equatorial and axial Co--O bond distances exhibit opposite temperature dependences; for example, the Co--O$_e$ (Co--O$_a$) distance increases (decreases) from 20 to $\sim60~\mathrm{K}$, remains nearly constant up to 100~K, then decreases (increases) sharply up to 160 K, followed by a much weaker reduction (enhancement) at higher temperatures up to 300~K [see Figs.~\ref{11_EXAFS_fit}(d, f)]. In contrast, the DW factors for both paths show similar temperature dependences, with a distinct minimum around 60~K [see Figs.~\ref{11_EXAFS_fit}(e, g)]. The strength of the Co--Sr/La peaks in the $\chi(R)$ spectra exhibits a trend opposite to that of the Co--O path, i.e., an increase from 20 to $\sim 60~\mathrm{K}$, followed by a sharp reduction for $60 < T \leqslant 130~\mathrm{K}$, and finally a gradual decrease for $T \geqslant 160~\mathrm{K}$, as shown in Fig.~\ref{11_EXAFS_fit}(c) (see also Figs.~S7(d--f) and S7(h) in Ref.~\cite{SI}). The temperature dependences of the Co--Sr and Co--La bond distances are shown in Figs.~\ref{11_EXAFS_fit}(h) and \ref{11_EXAFS_fit}(j), respectively, and their common Debye--Waller (DW) factor and average bond length are displayed in Figs.~\ref{11_EXAFS_fit}(i) and \ref{11_EXAFS_fit}(k). As expected from the behavior of the $\chi(R)$ spectra, the static distances for the Co--Sr and Co--La paths, first decrease up to around $60~\mathrm{K}$, then increase at $T \sim 120~\mathrm{K}$, and remain nearly invariant at higher temperatures, which is likewise reflected in their average path length, shown in Fig.~\ref{11_EXAFS_fit}(k). In contrast, their common DW factor varies smoothly with temperature, showing only a weak minimum around $60~\mathrm{K}$ [see Fig.~\ref{11_EXAFS_fit}(i)]. Furthermore, both Co--O--Nb forward scattering and Co--O2 single scattering path lengths and their corresponding DW factors also display significant modifications at the same temperatures [Figs.~\ref{11_EXAFS_fit}(l, m) and \ref{11_EXAFS_fit}(n, o), respectively]. 

\section{\noindent ~Discussion}

The high-energy Raman modes P1--P4, as well as the low-energy modes S7, S14, and S15, exhibit changes in intensity, linewidth, and peak position around 60 K. In the magnetically ordered state, spin–phonon coupling leads to an abrupt renormalization of the phonon self-energy, i.e., the frequency and linewidth of the modes \cite{Granado_PRB_99, Laverdiere_PRB_06, Graham_PRB_22}. However, neutron diffraction, magnetization, and specific heat measurements indicate that antiferromagnetic ordering of the Co$^{2+}$ moments in SrLaCoNbO$_6$ sets in only below $\sim$15 K \cite{Bos_PRB_04, Kumar_PRB1_20, Kumar_PRB2_20}. EXAFS measurements clearly show a change in the local coordination environment around the Co atoms near 60~K. At the same time, temperature-dependent neutron diffraction measurements reveal that the overall monoclinic $P2_1/n$ crystal structure persists down to 4 K \cite{Bos_PRB_04}. Therefore, a temperature-induced subtle modification of the local crystal structure, such as tilting, rotation, and/or deformation of the (Co/Nb)O$_6$ octahedra, could be responsible for the observed changes in the Raman modes around 60 K. This effect is more pronounced for the high-energy P1–P4 modes, which correspond to in-phase or out-of-phase stretching vibrations of the oxygen atoms in the octahedra. In particular, the opposite temperature dependences of the Co--O$_e$ and Co--O$_a$ bond distances provide a direct structural basis for the anomalies observed in these high-energy Raman modes. 

A similar deviation in the phonon self-energy parameters, well above the antiferromagnetic transition temperature ($\sim 3T_{\rm N}$), was also observed in Eu$_2$ZnIrO$_6$ and attributed to broken spin-rotational symmetry, suggesting the presence of short-range magnetic correlations deep within the paramagnetic phase \cite{Singh_PRR1_20}. Such short-range magnetic correlations may contribute to the subtle structural changes and the corresponding Raman response observed well above the spin-solid phase in the present system. The competition among inequivalent 90$^\circ$ Co$^{2+}$–O–Nb$^{5+}$–O–Co$^{2+}$ superexchange pathways within the fcc Co sublattice generates magnetic frustration in SrLaCoNbO$_6$ \cite{Bos_PRB_04}, which is also reflected in the relatively small specific-heat anomaly compared with the value expected for a pseudospin-$1/2$ system \cite{Kumar_PRB2_20}. This frustration can give rise to short-range magnetic correlations in the sample above the well-defined long-range-ordered phase. For example, a magnetostrictive effect well above the magnetic ordering temperature has recently been observed in Na$_5$Co$_{15.5}$Te$_6$O$_{36}$ and attributed to short-range magnetic interactions \cite{Saha_PRB_23}. Likewise, sub picometer scale magnetostriction due to spin–lattice coupling has been reported in the multiferroic perovskite DyMnO$_3$ \cite{Narayanan_PRB_17}. However, no observable anomalies in either the magnetization \cite{Kumar_PRB1_20} or the specific-heat data \cite{Kumar_PRB2_20} of SrLaCoNbO$_6$ were detected around 60 K. Therefore, it is difficult to ascertain whether short-range magnetic correlations drive the observed structural changes in the sample through magnetostriction or whether these changes are purely structural in origin. In this context, temperature-dependent EXAFS measurements at the nonmagnetic Nb edge are highly desirable to further clarify the origin of the local structural modifications and the associated Raman response near 60 K. 

The low-energy Raman modes S4–S6 exhibit asymmetric Fano line shapes, indicating coupling of these discrete phonons to an underlying continuum. Notably, a sizable background continuum is evident in the low-energy Raman spectra of SrLaCoNbO$_6$ for both laser excitation energies, 514 nm and 633 nm (see Figs.~S8 and S9 of \cite{SI}), which rules out luminescence as its origin. The broad continuum persists even in the long-range magnetically ordered phase at 12~K, in contrast to the well-defined broad peaks typically observed for two-magnon excitations \cite{Choi_PRB_04, Lake_NM_05, Lemmens_PRL_2000}, which suggests that magnons are a less probable origin in the present case. Furthermore, the highly insulating nature of SrLaCoNbO$_6$ \cite{Kumar_PRB1_20} makes a continuum associated with itinerant electronic carriers unlikely as the origin of the Fano line shape in these low-energy modes \cite{Sandilands_PRL_15}. In addition to the possible magnetic correlations, localization-related processes, orbital/electronic correlations, and dynamic lattice fluctuations may contribute to the observed response. For example, orbital fluctuations were identified as the dominant contribution to the pronounced Fano asymmetries of the low-energy phonons in the Mott insulator LaTiO$_3$ \cite{Ulrich_PRL_15}. Moreover, the broad Raman background may also contain a contribution from site-dependent oxygen-ion polarizability, since variations in the local oxygen environment can induce local lattice deformations and thereby promote lattice-polaron-like behavior \cite{Bilz_PRB_87, Holder_JPCM_12}.

Recently, a Fano line shape of magnetic origin has been reported in several 5$d$ magnetic insulators, such as $\alpha$-Ru$_{1-x}$Ir$_x$Cl$_3$ for $x=0$ \cite{Nasu_NP_16, Do_PRL_20, Sandilands_PRL_15} and $x=0.2$ \cite{Do_PRL_20}, $\beta$- and $\gamma$-Li$_2$IrO$_3$ \cite{Glamazda_NatCom_16}, and A$_2$ZnIrO$_6$ (A = Gd \cite{Singh_PRB_21} and Sm \cite{Singh_PRR2_20}). In these systems, phonon anomalies have been discussed in terms of inelastic scattering mediated by fractionalized quantum excitations, such as Majorana fermions. These materials typically exhibit a broad low-energy Raman continuum and display proximate Kitaev spin-liquid behavior above the magnetically ordered ground state \cite{Nasu_NP_16, Do_PRL_20, Sandilands_PRL_15, Glamazda_NatCom_16, Singh_PRB_21, Singh_PRR2_20, Wang_NQM_20}. Therefore, the excitation-energy-independent background continuum and the Fano line shape of the low-energy Raman modes are consistent with the presence of additional low-energy correlations in SrLaCoNbO$_6$ above the magnetic ordering temperature. Magnetic fluctuations constitute one possible contribution to this response. In this context, the theoretical predictions of Liu \textit{et al.} \cite{Liu_PRB_18}, which propose the possible realization of bond-dependent interactions in certain $d^7$ Co$^{2+}$ systems, provide an interesting perspective. However, in SrLaCoNbO$_6$, the Co ions are coupled through Co--O--Nb--O--Co super-superexchange pathways rather than the conventional edge-sharing Co--O--Co geometry. Therefore, further theoretical studies, together with inelastic neutron scattering and/or muon spin relaxation measurements, would be required to determine whether Kitaev-type interactions or related fractionalized magnetic excitations are relevant in this system. 

Further, in SrLaCoNbO$_6$, additional phonon anomalies around 160 K are manifested as changes in the intensity, linewidth, and peak position of the high-energy P1–P4 modes, variations in the intensities of the low-energy modes, a modification of the Fano asymmetry parameter $q$, and the emergence of an additional low-energy peak near 35 cm$^{-1}$. The EXAFS measurements also reveal notable changes in several bond distances around the Co atoms and in the corresponding DW factors near this temperature, suggesting a structural contribution to the Raman anomalies. Interestingly, the zero-field-cooled (ZFC) and FC magnetization curves bifurcate and exhibit a hump in the magnetic susceptibility of SrLaCoNbO$_6$ in this temperature range \cite{Kumar_PRB1_20}. Because the DW factor reflects incoherent atomic displacements along the scattering paths, the observed variations in pair distances and DW factors near 160 K indicate significant modifications of the local structural environment, which may also contain a magnetoelastic contribution. Notably, no discernible Co $K$-edge XANES shift or significant spectral change is observed near 60~K or 160--180~K, making valence-driven structural instabilities unlikely to account for the observed anomalies. However, at present it is difficult to determine whether thermally assisted local structural modifications give rise to the change in magnetization behavior or, conversely, whether changes in magnetic interactions drive the structural response via magnetostriction.

\section{\noindent ~Conclusions}

In summary, our temperature-dependent Raman and Co $K$-edge EXAFS measurements reveal a pronounced evolution of the lattice dynamics and local structure of the B-site-ordered double perovskite SrLaCoNbO$_6$. Several Raman-active phonon modes deviate from conventional anharmonic behavior, with distinct changes around $\sim$60~K and again near 160--180~K, both well above the long-range antiferromagnetic ordering temperature, $T_{\rm N} \approx 15$~K. The EXAFS analysis provides a direct structural counterpart to these anomalies, revealing correlated changes in the local CoO$_6$ octahedra and the broader Co coordination environment. In particular, the distinct evolution of the equatorial and axial Co--O bond distances, together with corresponding changes in the Debye--Waller factors and other Co-centered scattering paths, demonstrates subtle but systematic rearrangements of the local structure over the same temperature ranges. The low-energy S4--S6 modes exhibit asymmetric line shapes that are well described by Fano profiles, indicating coupling of the discrete phonons to an underlying continuum. The possible microscopic origins of this continuum and the associated asymmetric line shapes, including magnetic, electronic, orbital, localization-related, and lattice fluctuation effects, are discussed. These observations show that the anomalous phonon response is closely linked to modifications of the local Co--O framework, while correlation effects may provide an additional contribution. Overall, our results establish a strong connection between lattice dynamics and local structural rearrangements in SrLaCoNbO$_6$, highlighting the central role of the CoO$_6$ environment in governing its temperature-dependent vibrational properties. 

  \section{\noindent ~Acknowledgments}

AK thanks the UGC, India for the fellowship. RSD gratefully acknowledges the Department of Science and Technology (DST), India, for support through Indo-Australia Early and Mid-career Researchers fellowship (IA/INDOAUST/F-19/2017/1887) and UNSW for hosting the visit. We also thank Ravi Kumar and S. N. Jha for help and support during the temperature dependent XAS measurements at RRCAT, Indore, India. C.U. thanks the Australian Research Council for support through Grant No. LE110100060. RSD also thanks SERB (now ANRF) for financial support through a core research grant (file no.: CRG/2020/003436).

  \section{\noindent ~Data Availability}
  
  The data that support the findings of this article are not publicly available. The data are available from the authors upon reasonable request.


\begin{thebibliography}{99}

\bibitem{Liu_PRL_20} H. Liu, J. Chaloupka, and G. Khaliullin, Kitaev spin liquid in 3$d$ transition metal compounds, Phys. Rev. Lett. {\bf 125}, 047201 (2020).

\bibitem{Sano_PRB_18} R. Sano, Y. Kato, and Y. Motome, Kitaev-Heisenberg Hamiltonian for high-spin d$^7$ Mott insulators, Phys. Rev. B {\bf 97}, 014408 (2018). 

\bibitem{Chen_PRB_21} W. Chen, X. Li, Z. Hu, Ze Hu, Li Yue, R. Sutarto, F. He, K. Iida, K. Kamazawa, W. Yu, X. Lin, and Y. Li, Spin-orbit phase behavior of Na$_2$Co$_2$TeO$_6$ at low temperatures, Phys. Rev. B {\bf 103}, L180404 (2021).

\bibitem{Shukla_PRB_18}R. Shukla and R. S. Dhaka, Anomalous magnetic and spin glass behavior in Nb substituted LaCo$_{1-x}$Nb$_x$O$_3$, Phys. Rev. B {\bf 97}, 024430 (2018). 

\bibitem{Shukla_PRB_23}R. Shukla and R. S. Dhaka, Evolution of complex magnetic phases and metal-insulator transition through Nb substitution in La$_{0.5}$Sr$_{0.5}$Co$_{1-x}$Nb$_x$O$_3$, Phys. Rev. B {\bf 107}, 165108 (2023). 

\bibitem{Vasala_SSC_15} S.~Vasala and M.~Karppinen, A$_2$B$'$B$''$O$_6$ perovskites: A review, Prog. Solid State Chem. \textbf{43}, 1 (2015).

\bibitem{GarciaHernandez_PRL_01} M. Garc\'ia-Hern\'andez, J. L. Mart\'inez, M. J. Mart\'inez-Lope, M. T. Casais, and J. A. Alonso, Finding universal correlations between cationic disorder and low field magnetoresistance in FeMo double perovskite series, Phys. Rev. Lett. \textbf{86}, 2443 (2001).

\bibitem{Kumar_PRB1_20} A. Kumar and R. S. Dhaka, Unraveling magnetic interactions and the spin state in insulating Sr$_{2-x}$La$_x$CoNbO$_6$, Phys. Rev. B {\bf101}, 094434 (2020).

\bibitem{deVries_PRL_10} M. A. de Vries, A. C. McLaughlin, and J.-W. G. Bos, Valence bond glass on an fcc lattice in the double perovskite Ba$_2$YMoO$_6$, Phys. Rev. Lett. \textbf{104}, 177202 (2010).

\bibitem{Bos_PRB_04} J.-W. G. Bos and J. P. Attfield, Magnetic frustration in (LaA)CoNbO$_6$ (A = Ca, Sr, and Ba) double perovskites, Phys. Rev. B {\bf 70}, 174434 (2004). 

\bibitem{Aharen_PRB_09} T. Aharen, J. E. Greedan, F. Ning, T. Imai, V. Michaelis, S. Kroeker, H. Zhou, C. R. Wiebe, and L. M. D. Cranswick, Magnetic properties of the $S=3/2$ geometrically frustrated double perovskites La$_2$LiRuO$_6$ and Ba$_2$YRuO$_6$, Phys. Rev. B \textbf{80}, 134423 (2009).

\bibitem{Nilsen_PRB_15} G. J. Nilsen, C. M. Thompson, G. Ehlers, C. A. Marjerrison, and J. E. Greedan, Diffuse magnetic neutron scattering in the highly frustrated double perovskite Ba$_2$YRuO$_6$, Phys. Rev. B \textbf{91}, 054415 (2015).

\bibitem{Singh_PRR2_20} B. Singh, M. Vogl, S. Wurmehl, S. Aswartham,  B. B\"uchner, and P.  Kumar, Kitaev magnetism and fractionalized excitations in double perovskite Sm$_2$ZnIrO$_6$, Phys. Rev. Res. {\bf 2}, 013040 (2020).

\bibitem{Singh_PRB_21} B. Singh, D. Kumar, V. Kumar, M. Vogl, S. Wurmehl, S. Aswartham,  B. B\"uchner, and P.  Kumar, Fractional spin fluctuations and quantum liquid signature in  Gd$_2$ZnIrO$_6$, Phys. Rev. B {\bf 104}, 134402 (2021).

\bibitem{Glamazda_NatCom_16} A. Glamazda, P. Lemmens, S. -H. Do, Y. S. Choi, and K. -Y. Choi, Raman spectroscopic signature of fractionalized excitations in the harmonic-honeycomb iridates $\beta$- and $\gamma$-Li$_2$IrO$_3$, Nat. Commun. {\bf 7}, 12286 (2016).

\bibitem{Kimchi_PRB_14} I. Kimchi and A. Vishwanath, Kitaev-Heisenberg models for iridates on the triangular, hyperkagome, kagome, fcc, and pyrochlore lattices, Phys. Rev. B {\bf 89}, 014414 (2014).

\bibitem{Cook_PRBR_15} A. M. Cook, S. Matern, C. Hickey, A. A. Aczel, and A. Paramekanti, Spin-orbit coupled $j_{\rm eff}= $1/2 iridium moments on the geometrically frustrated fcc lattice, Phys. Rev. B {\bf 92}, 020417(R) (2015).

\bibitem{Aczel_PRB_19} A. A. Aczel, J. P. Clancy, Q. Chen, H. D. Zhou, D. Reigi-Plessis, G. J. MacDougall, J. P. C. Ruff, M. H. Upton, Z. Islam, T. J. Williams, S. Calder, and J.-Q. Yan, Revisiting the Kitaev material candidacy of  Ir$^{ 4+}$ double perovskite iridates, Phys. Rev. B {\bf 99}, 134417 (2019).
 
 \bibitem{Do_PRL_20} S.-H. Do, C. H. Lee, T. Kihara, Y. S. Choi, S. Yoon, K. Kim, H. Cheong, W.-T. Chen, F. Chou, H. Nojiri, and K.-Y. Choi, Randomly Hopping Majorana Fermions in the Diluted Kitaev System $\alpha$-Ru$_{0.8}$Ir$_{0.2}$Cl$_3$, Phys. Rev. Lett. {\bf 124}, 047204 (2020).
 
  \bibitem{Mabbs_book_73} F. E. Mabbs and D. J. Machin, Magnetism and transition metal complexes (Chapman and Hall, London, 1973).

\bibitem{Lloret_ICA_08} F. Lloret, M. Julve, J. Cano, R. Ruiz-Garc\'ia, and E. Pardo, Magnetic properties of six-coordinated high-spin Co(II) complexes: theoretical background and its application, Inorganica Chim. Acta  {\bf 361}, 3432 (2008).

\bibitem{Liu_PRB_18} H. Liu and G. Khaliullin, Pseudospin exchange interactions in ${d}^{7}$ cobalt compounds: Possible realization of the Kitaev model, Phys. Rev. B {\bf 97}, 014407 (2018). 

\bibitem{Kumar_PRB2_20} A. Kumar, B. Schwarz, H. Ehrenberg, and R. S. Dhaka, Evidence of discrete energy states and cluster-glass behavior in Sr$_{2-x}$La$_x$CoNbO$_6$, Phys. Rev. B {\bf102}, 184414 (2020).

\bibitem{Kumar_PRB_22} A. Kumar, R. Shukla, R. Kumar, R. J. Choudhary, S. N. Jha, and R. S. Dhaka, Probing the electronic and local structure of  Sr$_{2-x}$La$_x$CoNbO$_6$ using near-edge and extended x-ray absorption fine structures, Phys. Rev. B {\bf105}, 245155 (2022).

\bibitem{Cottam_book_86} M. G. Cottam and D. J. Lockwood, Light Scattering in Magnetic Solids (Wiley, New York, 1986).

\bibitem{Ulrich_PRL_06} C. Ulrich, A. G\"ossling, M. Gr\"uninger, M. Guennou, H. Roth, M. Cwik, T. Lorenz, G. Khaliullin, and B. Keimer, Raman Scattering in the Mott Insulators LaTiO$_3$ and YTiO$_3$: Evidence for Orbital Excitations, Phys. Rev. Lett. {\bf 97}, 157401 (2006).

\bibitem{Ulrich_PRL_09} C. Ulrich, L. J. P. Ament, G. Ghiringhelli, L. Braicovich, M. M. Sala, N. Pezzotta, T. Schmitt, G. Khaliullin, J. van den Brink, H. Roth, T. Lorenz, and B. Keimer, Momentum dependence of orbital excitations in Mott-insulating titanates, Phys. Rev. Lett. {\bf 103}, 107205 (2009).

\bibitem{Ulrich_PRL_15} C. Ulrich, G. Khaliullin, M. Guennou, H. Roth, T. Lorenz, and B. Keimer, Spin-Orbital Excitation Continuum and Anomalous Electron-Phonon Interaction in the Mott Insulator LaTiO$_3$, Phys. Rev. Lett. {\bf 115}, 156403 (2015).

\bibitem{Kumar_PRB_12}  P. Kumar, A. Bera, D. V. S. Muthu, S. N. Shirodkar, R. Saha, A. Shireen, A. Sundaresan, U. V. Waghmare, A. K. Sood, and C. N. R. Rao, Coupled phonons, magnetic excitations, and ferroelectricity in AlFeO$_3$: Raman and first-principles studies, Phys. Rev. B {\bf 85}, 134449 (2012).

\bibitem{Rovillain_PRB_10} P. Rovillain, M. Cazayous, Y. Gallais, A. Sacuto, M. -A. Measson, and H. Sakata, Magnetoelectric excitations in multiferroic TbMnO$_3$ by Raman scattering, Phys. Rev. B {\bf 81}, 054428 (2010).

\bibitem{Sohn_PRL_17} C. H. Sohn, C. H. Kim, L. J. Sandilands, N. T. M. Hien, S. Y. Kim, H. J. Park, K. W. Kim, S. J. Moon, J. Yamaura, Z. Hiroi, and T. W. Noh, Strong spin-phonon coupling mediated by single ion anisotropy in the all-in-all-out pyrochlore magnet Cd$_2$Os$_2$O$_7$, Phys. Rev. Lett. \textbf{118}, 117201 (2017).

\bibitem{Laverdiere_PRB_06} J. Laverdi\`ere, S. Jandl, A. A. Mukhin, V. Yu. Ivanov, V. G. Ivanov, and M. N. Iliev, Spin-phonon coupling in orthorhombic $R$MnO$_3$ ($R=$ Pr, Nd, Sm, Eu, Gd, Tb, Dy, Ho, Y): A Raman study, Phys. Rev. B \textbf{73}, 214301 (2006).

\bibitem{Singh_PRR1_20} B. Singh, M. Vogl, S. Wurmehl, S. Aswartham,  B. B\"uchner, and P.  Kumar, Coupling of lattice, spin, and intraconfigurational excitations of Eu$^{3+ }$ in Eu$_2$ZnIrO$_6$, Phys. Rev. Res. {\bf 2}, 043179 (2020).

\bibitem{Kamba_PRB_14} S. Kamba, V. Goian, V. Skoromets, J. Hejtm\'anek, V. Bovtun, M. Kempa, F. Borodavka, P. Van\v{e}k, A. A. Belik, J. H. Lee, O. Pacherov\'a, and K. M. Rabe, Strong spin-phonon coupling in infrared and Raman spectra of SrMnO$_3$, Phys. Rev. B \textbf{89}, 064308 (2014).

\bibitem{Iliev_APL_07} M. N. Iliev, H. Guo, and A. Gupta, Raman spectroscopy evidence of strong spin-phonon coupling in epitaxial thin films of the double perovskite La$_2$NiMnO$_6$, Appl. Phys. Lett. \textbf{90}, 151914 (2007).

\bibitem{Chakkar_PRB_25} A. G. Chakkar, C. B. Auti, D. Kumar, N. Jana, K. Pal, and P. Kumar, Raman signature of multiple phase transitions and quasiparticle excitations in the putative Kitaev spin liquid candidate Na$_2$Co$_2$TeO$_6$, Phys. Rev. B \textbf{112}, 144417 (2025).

\bibitem{Sandilands_PRL_15}  L. J. Sandilands, Y. Tian, K. W. Plumb, Y.-J. Kim, and K. S. Burch, Scattering Continuum and Possible Fractionalized Excitations in $\alpha$-RuCl$_3$, Phys. Rev. Lett. {\bf 114}, 147201 (2015).

\bibitem{Nasu_NP_16} J. Nasu, J. Knolle, D. L. Kovrizhin, Y. Motome, and R. Moessner, Fermionic response from fractionalization in an insulating two-dimensional magnet, Nature Phys. {\bf 12}, 912 (2016). 

\bibitem{Rodrigues_JMCC_22} J. E. Rodrigues, A. D. Rosa, J. L\'opez-S\'anchez, E. Sebastiani-Tofano, N. M. Nemes, J. L.  Martínez, J. A. Alonso, O. Mathon,  EXAFS evidence for the spin-phonon coupling in monoclinic PrNiO$_3$ nickelate perovskite, J. Mater. Chem. C {\bf 11}, 462 (2023).

\bibitem{Bridges_PRB_07} Y. Jiang, F. Bridges, L. Downward, and J. J. Neumeier, Relationship between macroscopic physical properties and local distortions of low-doping La$_{1-x}$Ca$_x$MnO$_3$: An EXAFS study, Phys. Rev. B {\bf 76}, 224428 (2007).

\bibitem{Panchwanee_PCM_19} A. Panchwanee, I. Schiesaro, S. Mobilio, S. S. K. Reddy, C. Meneghini, E. Welter, and V. R. Reddy, An evidence of local structural disorder across spin-reorientation transition in DyFeO$_3$: an extended x-ray absorption fine structure (EXAFS) study, J. Phys.: Condens. Matter {\bf 31}, 345403  (2019). 

\bibitem{SI} See the supplementary information at http://link.aps.org/supplemental/xxxxx  for the raw Raman data and details of the background continuum in the Raman spectrum, also includes \cite{Rousochatzakis_PRB_19}. 

\bibitem{Andrews_DT_15} R. L. Andrews, A. M. Heyns, and P. M. Woodward, Raman studies of A$_2$MWO$_6$ tungstate double perovskites, Dalton Trans. {\bf 44}, 10700 (2015).

\bibitem{Iliev_PRB_07} M. N. Iliev, M. V. Abrashev, A. P. Litvinchuk, V. G. Hadjiev, H. Guo, and A. Gupta, Raman spectroscopy of ordered double perovskite La$_2$CoMnO$_6$ thin films, Phys. Rev. B {\bf 75}, 104118 (2007).

\bibitem{Ayala_JAP_07} A. P. Ayala, I. Guedes, E. N. Silva, M. S. Augsburger, M. del C. Viola, and J. C. Pedregosa, Raman investigation of A$_2$CoBO$_6$ (A = Sr and Ca, B = Te and W) double perovskites, J. Appl. Phys. {\bf 101}, 123511 (2007).

\bibitem{Singh_PRM_20} B. Singh, M. Vogl, S. Wurmehl, S. Aswartham,  B. B\"uchner, and P.  Kumar, Kramers doublets, phonons, crystal-field excitations, and their coupling in Nd$_2$ZnIrO$_6$, Phys. Rev. Res. {\bf 2}, 023162 (2020).

\bibitem{Ajay_JAP20} A. Kumar, R. Shukla, A. Pandey, S. Dalal, M. Miryala, K. Ueno, M. Murakami, and R. S. Dhaka, Structural, transport, optical, and electronic properties of Sr$_2$CoNbO$_6$ thin films, J. Appl. Phys. {\bf 128}, 025303 (2020).

\bibitem{Glamazda_PRB_17} A. Glamazda, P. Lemmens, S.-H. Do, Y. S. Kwon, and K.-Y. Choi, Relation between Kitaev magnetism and structure in $\alpha$-RuCl$_3$, Phys. Rev. B {\bf 95}, 174429 (2017).

\bibitem{Fano_PR_124} U. Fano, Effects of Configuration Interaction on Intensities and Phase Shifts, Phys. Rev. {\bf 124}, 1866 (1961). 

\bibitem{Balkanski_PRB_83} M. Balkanski, R. F. Wallis, and E. Haro, Anharmonic effects in light scattering due to optical phonons in silicon, Phys. Rev. B {\bf 28}, 1928 (1983). 

\bibitem{Shukla_PRB22} R. Shukla, C. Ulrich, and R. S. Dhaka, Investigation of lattice dynamics, magnetism and electronic transport in $\beta$-Na$_{0.33}$V$_2$O$_5$, Phys. Rev. B {\bf 106}, 125148 (2022).

\bibitem{Menahem_PRM_23}  M. Menahem, N. Benshalom, M. Asher, S. Aharon, R. Korobko, O. Hellman, and O. Yaffe, Disorder origin of Raman scattering in perovskite single crystals, Phys. Rev. Materials {\bf 7}, 044602 (2023).

\bibitem{Michel_PRB_78}  K. H. Michel, J. Naudts, and B. De Raedt, Soft modes and central peak in orientationally disordered crystals, Phys. Rev. B {\bf 18}, 648 (1978).

\bibitem{Sokoloff_PRB_90} J. P. Sokoloff, L. L. Chase, and L. A. Boatner, Low-frequency relaxation modes and structural disorder in KTa$_{1-x}$Nb$_x$O$_3$, Phys. Rev. B {\bf 41}, 2398 (1990). 

\bibitem{Halperin_PRB_76} B. I. Halperin and C. M. Varma, Defects and the central peak near structural phase transitions, Phys. Rev. B {\bf 14}, 4030 (1976). 

\bibitem{Milkus_PRB_16} R. Milkus and A. Zaccone, Local inversion-symmetry breaking controls the boson peak in glasses and crystals, Phys. Rev. B {\bf 93}, 094204 (2016).

\bibitem{Choi_PRB_11} K.-Y. Choi, P. Lemmens, and H. Berger, Critical spin dynamics of the S = 1/2 spin chain compound CuSe$_2$O$_5$, Phys. Rev. B {\bf 83}, 174413 (2011). 

\bibitem{Wulferding_PRB_10} D. Wulferding, P. Lemmens, P. Scheib, J. Röder, P. Mendels, S. Chu, T. Han, and Y. S. Lee, Interplay of thermal and quantum spin fluctuations in the kagome lattice compound herbertsmithite, Phys. Rev. B {\bf 82}, 144412 (2010).

\bibitem{Yamada_PRB_94} I. Yamada and H. Onda, Light scattering from magnetic-energy fluctuations in the one-dimensional Heisenberg antiferromagnet KCuF$_3$, Phys. Rev. B {\bf 49}, 1048 (1994).

\bibitem{Shannon_AC_79} R. D. Shannon, Revised effective ionic radii and systematic studies of interatomic distances in halides and chalcogenides, Acta Crystallogr. Sect. {\bf A 32}, 751 (1976). 

\bibitem{Kuroe_PRB_97}  H. Kuroe, J. Sasaki, T. Sekine, N. Koide, Y. Sasago, K. Uchinokura, and M. Hase, Spin fluctuations in CuGeO$_3$ probed by light scattering, Phys. Rev. B {\bf 55}, 409 (1997). 

\bibitem{Kumar_JPCL_22} A. Kumar, A. Jain, S. M. Yusuf, and R. S. Dhaka, Observation of anisotropic thermal expansion and the Jahn-Teller effect in double perovskites Sr$_{2-x}$La$_x$CoNbO$_6$ using neutron diffraction, J. Phys. Chem. Lett. {\bf 13 }, 3023 (2022).  

\bibitem{Ajay_PRB24} A. Kumar, B. Schwarz, and R. S. Dhaka, Correlation between the exchange bias effect and antisite disorder in Sr$_{2-x}$La$_x$CoNbO$_6$ ($x=$ 0, 0.2 ), Phys. Rev. B {\bf 109}, 104434 (2024).
 
 \bibitem{Granado_PRB_99} E. Granado, A. Garc\'ia, J. A. Sanjurjo, C. Rettori, I. Torriani, F. Prado, R. D. S\'anchez, A. Caneiro, and S. B. Oseroff, Magnetic ordering effects in the Raman spectra of La$_{1-x}$Mn$_{1-x}$O$_3$, Phys. Rev. B {\bf 60}, 11879 (1999). 

\bibitem{Graham_PRB_22} P. J. Graham, P. Rovillain, M. Bartkowiak, E. Pomjakushina, K. Conder, M. Kenzelmann, and C. Ulrich, Spin-phonon and magnetoelectric coupling in oxygen-isotope substituted TbMnO$_3$ investigated by Raman scattering, Phys. Rev. B {\bf 105}, 174438 (2022).

\bibitem{Saha_PRB_23} R. A. Saha, J. Sannigrahi, I. Carlomagno, S. D. Kaushik, C. Meneghini, M. Itoh, V. Siruguri, and S. Ray, Short-range magnetic correlation, metamagnetism, and coincident dielectric anomaly in Na$_5$Co$_{15.5}$Te$_6$O$_{36}$, Phys. Rev. B {\bf 107}, 155105 (2023).

\bibitem{Narayanan_PRB_17} N. Narayanan, P. J. Graham, N. Reynolds, F. Li, P. Rovillain, J. Hester, J. Kimpton, M. Yethiraj, G. J. McIntyre, W. D. Hutchison, and C. Ulrich, Subpicometer-scale atomic displacements and magnetic properties in the oxygen-isotope substituted multiferroic DyMnO$_3$, Phys. Rev. B {\bf 95}, 075154 (2017). 

\bibitem{Choi_PRB_04} K.-Y. Choi, S. A. Zvyagin, G. Cao, and P. Lemmens, Coexistence of  dimerization and long-range magnetic order in the frustrated spin-chain system LiCu$_2$O$_2$: Inelastic light scattering study, Phys. Rev. B {\bf 69}, 104421 (2004).

\bibitem{Lake_NM_05} B. Lake, D. A. Tennant, C. D. Frost, and S. E. Nagler, Quantum criticality and universal scaling of a quantum antiferromagnet, Nat. Mater. {\bf 4}, 329 (2005).

\bibitem{Lemmens_PRL_2000} P. Lemmens, M. Grove, M. Fischer, G. Güntherodt, V. N. Kotov, H. Kageyama, K. Onizuka, and Y. Ueda, Collective Singlet Excitations and Evolution of Raman Spectral Weights in the 2D Spin Dimer Compound SrCu$_2$(BO$_3$)$_2$, Phys. Rev. Lett. {\bf 85}, 2605 (2000). 

\bibitem{Bilz_PRB_87} H. Bilz, G. Benedek, and A. Bussmann-Holder, Theory of ferroelectricity: The polarizability model, Phys. Rev. B {\bf 35}, 4840 (1987).

\bibitem{Holder_JPCM_12}A. Bussmann-Holder, The polarizability model for ferroelectricity in perovskite oxide, J. Phys.: Condens. Matter {\bf 24}, 273202 (2012).

\bibitem{Wang_NQM_20} Y. Wang, G. B. Osterhoudt, Y. Tian, P. Lampen-Kelley, A. Banerjee, T. Goldstein, J. Yan, J. Knolle, H. Ji, R. J. Cava, J. Nasu, Y. Motome, S. E. Nagler, D. Mandrus, and K. S. Burch, The range of non-Kitaev terms and fractional particles in $\alpha$-RuCl$_3$, npj Quantum Materials {\bf 5}, 14 (2020).

\bibitem{Rousochatzakis_PRB_19} I. Rousochatzakis, S. Kourtis, J. Knolle, R. Moessner, and N. B. Perkins, Quantum spin liquid at finite temperature: Proximate dynamics and persistent typicality, Phys. Rev. B {\bf 100}, 045117 (2019).

\end{thebibliography}
\end{document}